\input phyzzx
\overfullrule 0pt
\catcode `\@=11
\def\PH@SR@V{\doubl@true \baselineskip=18.1pt plus 0.1pt minus 0.1pt
             \parskip= 3pt plus 2pt minus 1pt }
\def\PHYSREV{\papers\PhysRevtrue\PH@SR@V}
\let\physrev=\PHYSREV
\def\FTUV{\address{Departamento de F\'{\i}sica Te\'{o}rica and IFIC,\break
      Centro Mixto Univ. de Valencia-CSIC\break 
	46100-Burjassot (Valencia), Spain}}
\VOFFSET=33pt
\papersize
\physrev
\font\black=msbm10 scaled\magstep1
\def\title#1{\vskip\frontpageskip \titlestyle{\seventeenbf #1}
 \vskip\headskip }
\def\abstract{\par\dimen@=\prevdepth \hrule height\z@ \prevdepth=\dimen@
   \vskip\frontpageskip\centerline{\twelverm ABSTRACT}\vskip\headskip }
\def\refmark#1{[#1]}
\def\NAME{\author{J.A. de Azc\'{a}rraga
\footnote{\tenrm\dag}{E-mail: azcarrag@evalvx.ific.uv.es}, 
A. M. Perelomov
\footnote{\tenrm\star}
{On leave of absence from Institute for Theoretical and
Experimental Physics, 117259 Moscow, Russia.\quad
E-mail: perelomo@evalvx.ific.uv.es} and 
J.C. P\'{e}rez Bueno \footnote{\tenrm\ddag}{E-mail: pbueno@lie.ific.uv.es}}}
\catcode `\@=\active

\def \acting {({\Number {7}}.6)}
\def\ld{\mathop{\ldots}\limits}
\def\frac#1#2{{#1\over#2}}

\def\campo{{\cal X}}

\def\field #1{\hbox{{\black #1}}}

\def\pois#1#2{\{#1,#2\}}         
\def\dd#1{\frac{\partial}{\partial#1}}
\def\set#1{\{\,#1\,\}}         
\def\R{{\hbox{{\field R}}}} 
\def\C{{\hbox{{\field C}}}}

\def\hat{\widehat}
 
\def\g{{\cal G}}

\def\lbr{{[\![}}
\def\rbr{{]\!]}}
\REF\Na{Nambu, Y.: 
{\it Generalized Hamiltonian dynamics},
Phys. Rev. {\bf D7}, 2405--2412 (1973)}

\REF\BF{Bayen, F. and Flato, M.: 
{\it Remarks concerning Nambu's generalized mechanics}, 
Phys. Rev. {\bf D11}, 3049--3053 (1975)}

\REF\MS{Mukunda, N. and Sudarshan E.: 
{\it Relation between Nambu and Hamiltonian mechanics}, 
Phys. Rev. {\bf D13}, 2846--2850 (1976)}

\REF\Hir{Hirayama, H.: 
{\it Realization of Nambu mechanics: A particle 
interacting with an $SU(2)$ monopole}, 
Phys. Rev. {\bf D16}, 530--532 (1977)}

\REF\SV{Sahoo, V. and Valsakumar, M. C.:
{\it Non-existence of quantum Nambu-mechanics},
Mod. Phys. Lett. {\bf A29}, 2727-2732 (1994)}

\REF\Ta{Takhtajan, L.:
{\it On foundations of the generalized Nambu mechanics},
Commun. Math. Phys. {\bf 160}, 295--315 (1994)}

\REF\Cha{Chatterjee, R.: {\it Dynamical symmetries and Nambu mechanics}, Lett. 
Math. Phys. {\bf 36}, 117-126 (1996)}

\REF\Lich{Lichnerowicz, A.:
{\it Les vari\'et\'es de Poisson et leurs alg\`ebres de Lie associ\'ees},
J. Diff. Geom. {\bf 12}, 253-300 (1977)} 

\REF\We{Weinstein, A.:
{\it The local structure of Poisson manifolds},
J. Diff. Geom. {\bf 18}, 523--557 (1983)}

\REF\BFFLS{Bayen, F.; Flato, M.; Fronsdal, C.; Lichnerowicz, A. 
and Sternheimer, D.: 
{\it Deformation theory and quantization},
Ann. Phys. {\bf 111}, 61--151 (1978)}

\REF\APPBJPA{de Azc\'arraga, J. A.; Perelomov, A. M. and P\'erez Bueno, J. C.:
{\it New Generalized Poisson Structures}, 
J. Phys. {\bf A29}, L151-157 (1996)}

\REF\Tu{Tulczyjev, W.M.: 
{\it Poisson brackets and canonical manifolds}, 
Bull. Acad. Pol. Sci. (Math. and Astronomy) {\bf 22}, 931--934 (1974)}

\REF\Sc{Schouten, J.A.:
{\it Ueber Differentialkomitanten zweir kontravarianter Gr\"oszen}, 
Proc. Kon. Ned. Akad. Wet. Amsterdam {\bf 43}, 449-452 (1940)}

\REF\Ni{Nijenhuis, A.: 
{\it Jacobi-type identities for bilinear differential concomitants 
of certain tensor fields}, 
Indag. Math. {\bf 17}, 390-403 (1955)}

\REF\Li{Lie, S.:
{\it Begr\"undung einer Invariantentheorie der Ber\"uhrungs Transformationen},
Math. Ann. {\bf 8}, 214--303 (1874/75)}

\REF\Lie{Lie, S. and Engel, F.: 
{\it Theorie der Transformationsgruppen I--III}, 
Teubner (1888) (Chelsea, 1970)}

\REF\GMP{Grabowski, J.; Marmo, G. and Perelomov, A. M.:
{\it Poisson structures: towards a classification},
Mod. Phys. Lett. {\bf A18}, 1719--1733 (1993)}

\REF\CIMP{Cari\~ nena, J.; Ibort, A.; Marmo, G. and Perelomov A. M.:
{\it On the geometry of Lie algebras and Poisson tensors},
J. Phys. {\bf A27}, 7425--7449 (1994)}

\REF\AP{Alekseevsky, D.V. and Perelomov, A.M.:
{\it Poisson brackets on Lie algebras} 	   
Preprint ESI 247 (1995), to appear in J. Geom. Phys.}

\REF\Kir{Kirillov, A. A.:
{\it Local Lie algebras},
Russian Math. Surveys {\bf 31}, 55-75 (Uspekhi Math. Nauk. {\bf 31}, 57-76) 
(1976)}

\REF\Koszul{Koszul, J. L.: 
{\it Crochet de Schouten-Nijenhuis et cohomologie}, 
Ast\'erisque (hors s\'erie) 257-271 (1985)}

\REF\CNS{Corwin, L.; Ne'eman, Y. and Sternberg, S.:
{\it Graded Lie algebras in mathematics and physics (Bose-Fermi symmetry)},
Rev. Mod. Phys. {\bf 47}, 573-603 (1975)}

\REF\Po{Poisson, S.:
{\it M\'emoire sur la variation des constantes arbitraires dans les 
questions de m\'echanique},
J. \'Ecole Polytec. {\bf 8}, 266--344 (1809)}

\REF\AG{Alekseevsky, D. and Guha, P:
{\it On decomposability of Nambu-Poisson tensor}, Bonn Inst. f\"ur Math. 
preprint MPI/96--9 (1996)}

\REF\ChaTak{Chatterjee, R. and Takhtajan, L.:
{\it Aspects of classical and quantum Nambu mechanics}, Lett. Math. Phys. {\bf 
37}, 475-482 (1996)}

\REF\CE{Chevalley, C. and Eilenberg, S.: {\it Cohomology theory of
Lie groups and Lie algebras}, Trans. Am. Math. Soc. {\bf 63}, 85--124 
(1948)} 

\REF\AI{de Azc\'arraga, J. A. and Izquierdo, J.M.: {\em Lie algebras, Lie 
groups, cohomology and some applications in physics}, Camb. Univ. Press 
(1995)}

\REF\DFST{Dito, G.; Flato, M.; Sternheimer, D. and Takhtajan, L.:
{\it Deformation Quantization and Nambu mechanics}, hep-th/9602016, to appear 
in Commun. Math. Phys.}

\REF\HIGHER{de Azc\'arraga, J. A. and P\'erez Bueno, J. C.:
{\it Higher-order simple Lie algebras}
FTUV/96-23 , IFIC/96-25 (hep-th/9605213)}

\REF\LAST{Lada, T. and Stasheff, J.:
{\it Introduction to SH Lie algebras for physicists},
Int. J. Theor. Phys. {\bf 32}, 1087-1103 (1993)}

\REF\WZ{Witten, E. and Zwiebach, B:
{\it Algebraic structures and differential geometry in two-dimensional string 
theory},
Nucl. Phys. {\bf B377}, 55-112 (1992)}

\REF\ZIE{Zwiebach, B.: 
{\it Closed string theory: quantum action and the Batalin-Vilkoviski master 
equation},
Nucl. Phys. {\bf B390}, 33-152 (1993)}

\REF\AIPP{de Azc\'arraga, J. A.; Izquierdo, J. M.; Perelomov, A. M. and P\'erez
Bueno, J. C.:
{\it Super-Poisson structures and their generalization}, in preparation}
\nopubblock
\titlepage
\rightline{FTUV/96--15}
\rightline{IFIC/96--17}
\title{The Schouten--Nijenhuis bracket, cohomology and
generalized Poisson structures}
\NAME
\FTUV
\abstract
Newly introduced generalized Poisson structures 
based on suitable skew--sym\-metric 
contravariant tensors of even order are discussed in terms of 
the Schouten-Nijenhuis bracket. 
The associated `Jacobi identities' are expressed as conditions on these tensors, 
the cohomological contents of which is given. 
In particular, we determine the linear generalized Poisson structures which 
can be constructed on the dual spaces of simple Lie algebras.
\endpage

\chapter{Introduction}
In 1973 Nambu \refmark{\Na} proposed a generalization of the 
standard classical Hamiltonian mechanics based on a three-dimensional 
`phase space' spanned by a canonical triplet of dynamical variables and on 
two `Hamiltonians'.
His approach was later discussed by Bayen and Flato \refmark{\BF} and \eg\ in 
\refmark{\MS,\Hir,\SV}.
Recently, a higher order extension of 
Nambu's approach, involving $(n-1)$ Hamiltonians, 
was proposed by Takhtajan \refmark{\Ta} 
(see \refmark{\Cha} for applications).
This approach, which includes Nambu's mechanics as a particular case, has the 
property that the time derivative is a derivation of the $n$-th order Poisson 
bracket (PB) because the expression of this fact, which involves $(n+1)$ terms, 
is the same as the `fundamental identity' \refmark{\Ta} which generalizes 
the Jacobi identity of the ordinary $n=2$ case.
Closely related to Hamiltonian dynamics is the study of 
Poisson structures (PS) (see \refmark{\Lich,\We,\BFFLS}) on a 
manifold $M$. 

Recently, a different generalization of PS has
been put forward \refmark{\APPBJPA}. In contrast with those of 
Nambu and Takhtajan, the dynamics is associated with generalized Poisson 
brackets (GPB) necessarily involving an even number 
of functions.
The aim of this paper is to discuss these new generalized Poisson structures 
(GPS) further and, in particular, to exhibit the cohomological contents 
of the examples provided (for the linear GPS) on the dual spaces of simple Lie 
algebras.
The key idea of the new GPS is the replacement of the skew-symmetric bivector 
$\Lambda$ defining the standard Poisson structure by appropriate 
skew-symmetric contravariant tensor fields of even order $\Lambda^{(2p)}$. 
For a standard $(p=1)$ PS, the property which guarantees the Jacobi identity 
for the PB of two functions on a Poisson manifold may be expressed 
\refmark{\Lich,\Tu} as $[\Lambda,\Lambda]=0$, where 
$\Lambda\equiv\Lambda^{(2)}$ is the bivector field which may be used to 
define the Poisson structure and $[\ ,\ ]$ is the Schouten--Nijenhuis 
bracket (SNB) \refmark{\Sc,\Ni}. 
Thus, a natural generalization of the standard PS may be found 
\refmark{\APPBJPA} using $\Lambda^{(2p)}$, and replacing 
Jacobi identity by the condition which follows from 
$[\Lambda^{(2p)},\Lambda^{(2p)}]=0$.
The vanishing of the SNB of $\Lambda^{(2p)}$ with itself  
generalizes the Jacobi identity in a geometrical way, 
different from \refmark{\Na,\Ta}: our GPB 
involve an {\it even} number of functions, whereas this number is 
arbitrary (three in \refmark{\Na}) in earlier extensions.

The geometrical content of the theory becomes specially apparent when 
the linear GPS on the duals $\g^*$ of simple Lie algebras
$\g$ are considered, since they automatically provide us with 
solutions of the generalized Jacobi identities (GJI) that our
Poisson structures must satisfy.
In fact, since the Jacobi identity or its generalizations constitute
the only essentially non-trivial ingredient of any PS
(the skew-symmetry and the Leibniz rule are easy to satisfy), 
it is important to have explicit examples which satisfy them.
In our linear GPS, 
the solution to the GJI has a cohomological character:
the different tensors
$\Lambda^{(2p)}$ that can be introduced are related to Lie 
algebra cocycles.

\chapter{Standard Poisson structures}

Let us recall for completeness some facts concerning standard PS.
Let $M$ be a manifold and ${\cal F}(M)$ be the associative algebra 
of smooth functions on $M$.
\medskip
\noindent
{\bf Definition 2.1 (PB)\quad}
A {\it Poisson bracket} $\{ \cdot ,\cdot \}$ on ${\cal F}(M)$ is a
bilinear mapping assigning to every pair of functions 
$f_1, f_2\in {\cal F}(M)$ a new
function $\{f_1, f_2\}\in {\cal F}(M)$, with 
the following conditions:

\noindent
a) skew-symmetry 
$$\{f_1, f_2\} = - \{f_2, f_1\}\;,\eqn\pri$$
b) Leibniz rule (derivation property)
$$\{f, gh\} = g\{f, h\} + \{f, g\}h \;,\eqn\ter$$
c) Jacobi identity (JI)
$${1\over 2}{\rm Alt} \{f_1,\{f_2,f_3\}\}\equiv
\{f_1, \{f_2, f_3\}\} + \{f_2, \{f_3,f_1\}\} +
\{f_3, \{f_1,f_2\}\}= 0\;;\eqn\sec$$
$M$ is then called a {\it Poisson manifold}.
Because of \pri, \sec\ 
the space ${\cal F}(M)$ endowed with the PB
$\{ \cdot ,\cdot \}$
becomes an (infinite-dimensional) Lie algebra.

Let $x^{j}$ be local coordinates on $U\subset M$ and consider a PB
of the form
$$
\{f(x), g(x)\} = \omega ^{jk}(x)\partial _{j}f\partial _{k}g\quad,\quad 
\partial _{j} = {\partial \over {\partial x^{j}}}\quad,\quad
j,k=1,\ldots, n={\rm dim}M 
\eqn\cua
$$
Then, $\omega^{ij}(x)$ defines a 
PB if
$\omega ^{ij}(x) = -\omega ^{ji}(x)$
(eq. \pri) and (eq. \sec) 
$$
\omega ^{jk}\partial _{k}\omega ^{lm} + \omega ^{lk}\partial _{k}\omega 
^{mj} + \omega ^{mk}\partial _{k}\omega ^{jl} = 0\;.
\eqn\sex
$$

The requirements \pri\ and \ter\ indicate 
that the PB may be given in terms of 
a skew-symmetric bivector field ({\it Poisson bivector}) 
$\Lambda\in\wedge ^2(M)$ which is uniquely defined. 
Locally,
$$
\Lambda = {1\over 2} 
\omega ^{jk}\partial _j\wedge \partial _k\quad.\eqn\nueva
$$
Condition \sex\ is taken into account by requiring 
$[\Lambda, \Lambda]=0$
\refmark{\Lich, \Tu} (Sec. 3). 
Then, $\Lambda$ 
defines a {\it Poisson structure} on $M$
and the PB is defined by 
$$
\{f,g\}=\Lambda(df,dg)\quad,\quad f,g\in{\cal F}(M)
\quad.\eqn\oneone
$$

\medskip
\noindent
{\bf Definition 2.2\quad}
Let $H(x)\in{\cal F}(M)$. Then the vector field $X_H=i_{dH}\Lambda$ 
(where 
$i_\alpha\Lambda(\beta):=\Lambda(\alpha,\beta)\,,\,\allowbreak\alpha,\beta$ 
one-forms), is called a
{\it Hamiltonian vector field} of $H$. 

>From the JI \sec\ easily follows that 
$$
[X_f,X_H]=X_{\{ f,H\} }\quad.
\eqn\sep
$$
Thus, the Hamiltonian vector fields form a Lie subalgebra of the Lie algebra 
$\campo(M)$ of all smooth vector fields on $M$.
Locally,
$$
X_{H}(x) = \omega ^{jk}(x) (\partial _{j}H(x))\partial _k\quad;
\quad
X_H.f =\{H,f\}\;.
\eqn\oct
$$
We recall that the tensor $\omega ^{jk}(x)$ appearing in \cua, \nueva\ 
does not need to
be nondegenerate; in particular, the dimension of a Poisson manifold 
$M$ may be odd.
Only when $\Lambda$ has constant rank $2q$ (is {\it regular}) 
and the codimension
(${\rm dim}M-2q$) of the manifold is zero, $\Lambda$ defines a
{\it symplectic structure}.

\chapter{Standard linear Poisson structures}

A particular class of Poisson structures is that defined 
on the duals $\g^*$ of Lie algebras $\g$.
The case of the linear Poisson structures was considered by Lie himself 
\refmark{\Li,\Lie}, and has been further investigated recently 
\refmark{\GMP,\CIMP,\AP}.
Let $\g$ be 
a real finite--dimensional Lie algebra $\g$ with Lie bracket
$[\, .,.]$. 
The natural identification $\g \cong
(\g^*)^*$, allows us to think of $\g$ as a subspace of linear functions 
of the ring of smooth functions ${\cal F}(\g^*)$.  Choosing a basis
$\set{e_i}_{i=1}^r$ of $\g$, 
$[e_i, e_j] =  C_{ij}^{k} e_k$, 
and identifying its elements with linear
coordinate functions $x_i$ on the dual space $\g^*$ by means of $x_i(x) =
\langle x, e_i\rangle$ for all $x\in \g^*$, the
fundamental PB on $\g^*$ may be defined in a natural way by taking 
$$
\pois{x_i}{x_j}_\g = C_{ij}^{k} x_k=\omega_{ij}(x) 
\quad,\quad i,j,k=1,\ldots, r={\rm dim}
\g\quad,
\eqn\IVv
$$
since the Jacobi identity for $C_{ij}^k$ implies that \sex\ is satisfied.
Intrinsically, the PB
$\pois . . _\g$ 
on 
${\cal F}(\g^*)$ 
is defined by
$$
\pois{f}{g}_\g (x) = \langle x,[df(x),dg(x)]\rangle\quad,\quad
f,g\in {\cal F}(\g^*),\ x\in \g^*\quad,
\eqn\IVvi
$$
where the one-forms in the bracket are regarded as linear mappings 
from $T_x(\g^*)\sim \g^*$ to $\R$ and hence as elements of $\g$. Locally, 
$$
[df(x),dg(x)]=e_kC_{ij}^k {\partial f\over\partial x_i}
{\partial g\over \partial x_j}
\quad,\quad
\{f,g\}_\g (x)=x_kC_{ij}^k{\partial f\over\partial x_i}
{\partial g\over \partial x_j}\quad.
\eqn\newformula
$$
The above PB
$\pois . . _\g$ 
(see \refmark{\Kir}) is commonly called a {\it Lie--Poisson bracket} 
and defines a {\it Lie-Poisson} structure on $\g^*$.
It is associated with the bivector field $\Lambda_\g$ on $\g^*$ 
locally written as 
$$
\Lambda_\g = {1\over 2}C_{ij}^{k} x_k \dd{x_i}\wedge \dd{x_j}
\equiv {1\over 2}\omega_{ij}\partial^i\wedge\partial^j
\eqn\IVvii
$$
(cf. \nueva), so that (cf. \oneone)
$\Lambda_\g (df,dg) = \pois{f}{g}_\g$.
Then,
$[\Lambda_\g,\Lambda_\g]_S=0$ (cf. \sex) leads to the Jacobi identity for $\g$,
which may be written as
$$
{1\over 2}
{\rm Alt}(C^\rho_{i_1 i_2}C^\sigma_{\rho i_3})\equiv
{1\over 2}
\epsilon^{j_1j_2j_3}_{i_1i_2i_3}C^\rho_{j_1 j_2}C^\sigma_{\rho j_3}=0\quad.
\eqn\VIa
$$
Notice also that the Poisson bracket of two polynomial functions on $\g^*$ is
again a polynomial function, so that the space ${\cal P}(\g^*)$ of all
polynomials on $\g^*$ is a Lie subalgebra.
 
Let 
$\beta$ be a closed one form on 
$\g^*$.
The associated vector field  
$$
X_\beta= i_\beta\Lambda_\g \quad,\eqn\IVx
$$
is an infinitesimal automorphism of $\Lambda_\g$ \ie,  
$$
L_{X_\beta} \Lambda_\g = 0\quad,\eqn\IVxa
$$
and 
$[X_f, X_g] = X_{\pois f g}$ (eq. \sep);
this is proved easily using that $L_{X_f}g = \pois f g $ and
$L_{X_f}\Lambda_\g= 0$.  
It follows from \IVvii\ that
the Hamiltonian vector fields 
$X_i=i_{dx_i}\Lambda_\g$
corresponding to the linear coordinate functions $x_i$, have the
expression (cf. \oct)
$$
X_i = C_{ij}^{k} x_k \dd{x_j}\quad,\quad i=1,\ldots,r={\rm dim}\g
\eqn\IVxii
$$
so that the Poisson bivector can be written as 
$$
\Lambda_\g = -{1\over 2}X_i \wedge \dd{x_i}\quad;
\eqn\IVxiii
$$
notice that this way of writing $\Lambda_\g$ is of course not unique.
Using the adjoint representation of 
$\g\,,\, (C_i)^k_{.j}=C^k_{ij}$
the Poisson bivector 
$\Lambda_\g$ may be rewritten as 
$$
\Lambda_\g = -{1\over 2} X_{C_i} \wedge \dd{x_i}\quad
\quad (X_{C_i}=x_k(C_i)^k_{.j}\dd{x_j})\quad;
\eqn\IVxvi
$$
the vector fields $X_{C_i}$ provide a realization of ${\rm ad}\g$
in terms of vector fields on $\g^*$.

\chapter{The Schouten--Nijenhuis bracket}

Let $\wedge(M)=\bigoplus^n_{j=0}\wedge^j(M)\ (\wedge^0={\cal F}(M) \,,\, 
n={\rm dim} M$), 
be the contravariant exterior algebra of skew-symmetric contravariant (\ie, 
tangent) tensor fields ({\it multivectors} or $j$-{\it vectors}) over $M$.
The Lie bracket of vector fields on $M$ may be uniquely extended to an 
$\R$-bilinear bracket on $\wedge(M)$, the SNB, in such way that $\wedge(M)$ 
becomes a graded superalgebra (see the Remark below). 
The SNB \refmark{\Sc,\Ni} is a bilinear mapping 
$\wedge^p (M)\times \wedge^q (M)\to \wedge^{p+q-1}(M)$. 
We start by defining the SNB for multivectors given by products of vector 
fields.

\medskip
\noindent
{\bf Definition 4.1\quad}
Let $X_1,\ldots, X_p, Y_1,\ldots, Y_q$ be vector fields over $M$. Then
$$
\eqalign{
[X_1\wedge...\wedge X_{p}\quad,&
\quad Y_{1}\wedge \ldots \wedge Y_{q}] =
\cr
= \sum (-1)^{t+s}X_{1}\wedge...\hat X_{s}...&
\wedge X_{p}
\wedge[X_{s} ,Y_{t}]\wedge Y_{1} \wedge ... \hat Y_{t} ...
\wedge Y_{q}\quad,
\cr}
\eqn\numerari
$$ 
where $[\ ,\ ]$ is the SNB and $\hat X$ stands for the omission of X.
It is easy to check that \numerari\ is equivalent to original definition
\refmark{\Sc,\Ni}. 

\medskip
\noindent
{\bf Theorem 4.1\quad}
Let $M=G$ be the group manifold of a Lie group, and let the above vector 
fields $X\,,\,Y$ be left-invariant [LI] 
(resp. right-invariant [RI]) vector fields on $G$.
Then, the SNB of LI (resp. RI) skew multivector fields is also LI (resp. RI).

\noindent
{\it Proof.\quad}
It suffices to recall that if $X$ is LI, $L_Z X=[Z,X]=0$ where $Z$ is the 
generator of the left translations, {\it q.e.d.}

\medskip
\noindent
{\bf Definition 4.2\quad}
Let $A\in \wedge^p(M)$ and $B\in \wedge^q(M)$, $p,q\leq n$, 
be the $p$- and $q$-vectors (multivectors of order $p$ and $q$ respectively) 
given in a local chart by
$$
A(x)={1\over p!}{A^{i_1\ldots i_p}(x)} \partial_{i_1}
\wedge\ldots\wedge \partial_{i_p}\quad,\quad
B(x)={1\over q!}{B^{j_1\ldots j_q}(x)} 
\partial_{j_1}\wedge\ldots\wedge \partial_{j_q}\quad.
\eqn\IIvii
$$
The SNB of $A$ and $B$ is the skew-symmetric contravariant tensor field 
$[A,B]\in \wedge^{p+q-1}(M)$
$$
\eqalign{[A,B]=&{1\over (p+q-1)!}
[A,B]^{k_1\ldots k_{p+q-1}}
\partial_{k_1}\wedge\ldots\wedge \partial_{k_{p+q-1}}\quad,\cr
[A,B]^{k_1\ldots k_{p+q-1}}=&{1\over (p-1)!q!}
\epsilon^{k_1\ldots k_{p+q-1}}_{i_1\ldots i_{p-1} j_1\ldots j_q}
A^{\nu i_1\ldots i_{p-1}}\partial_\nu B^{j_1\ldots j_q}\cr
&+{(-1)^p\over p!(q-1)!}
\epsilon^{k_1\ldots k_{p+q-1}}_{i_1\ldots i_{p} j_1\ldots j_{q-1}}
B^{\nu j_1\ldots j_{q-1}}\partial_\nu A^{i_1\ldots i_p}\quad,\cr}
\eqn\VIIii
$$
where $\epsilon$ is the antisymmetric Kronecker symbol,
$$
\epsilon^{i_1\ldots i_{p}}_{j_1\ldots j_{p}}=
{\rm det}\left(\matrix{\delta^{i_1}_{j_1} & \cdots & \delta^{i_1}_{j_p} \cr
\vdots & & \vdots \cr
\delta^{i_p}_{j_1} & \cdots & \delta^{i_p}_{j_p} \cr}\right)\quad.
\eqn\defep
$$

The SNB is graded-commutative, 
$$
[A,B]=(-1)^{pq}\,[B,A]\quad.
\eqn\IIv
$$
As a result, the SNB is identically zero if $A=B$ 
are of odd order (or even {\it degree};
${\rm degree}(A)\equiv{\rm order}(A)-1$).
It satisfies the graded Jacobi identity,
$$
(-1)^{pr}\,[[A,B],C]+(-1)^{qp}\,[[B,C],A]+(-1)^{rq}\,[[C,A],B]=0\quad,
\eqn\IIvi
$$
where $(p,q,r)$ denote the order of $(A,B,C)$ respectively
(thus, if $\Lambda$ is of even order and $[\Lambda,\Lambda]=0$ it follows from
\IIvi\ that $[\Lambda,[\Lambda,C]]=0$). 

Let $A\wedge B\in \wedge^{p+q}(M)$,
$$
\eqalign{
(A\wedge B)&={1\over (p+q)!}(A\wedge B)^{i_1\ldots i_{p+q}}
\partial_{i_1}\wedge\ldots\wedge\partial_{i_{p+q}}\quad,
\cr
(A\wedge B)&^{i_1\ldots i_{p+q}}={1\over p!q!}
\epsilon^{i_1\ldots i_{p+q}}_{j_1\ldots j_{p+q}}
A^{j_1\ldots j_{p}}B^{j_{p+1}\ldots j_{p+q}}\quad,
\cr}
\eqn\wedproduct
$$
and let $\alpha\in\wedge_{p+q-1}(M)$ be an arbitrary $(p+q-1)$-form,
$\alpha={1\over (p+q-1)!}
\alpha_{{i_1}\ldots i_{p+q-1}}dx^{i_1}\wedge\ldots\wedge dx^{i_{p+q-1}}$.
Then, the well known formula for one-forms and vector fields, 
$d\omega(X,Y)=L_X\omega(Y)-L_Y\omega(X)-i_{[X,Y]}\omega\,,$ generalizes to
$$
i_{A\wedge B}d\alpha=(-1)^{pq+q}i_A d(i_B \alpha)+(-1)^p i_B d(i_A\alpha)-
i_{[A,B]}\alpha
\quad,
\eqn\IIi
$$
where the contraction $i_A\alpha$ is the $(q-1)$-form
$$
i_A\alpha(\cdot)={1\over p!}\alpha(A,\cdot)\quad,\quad
i_A\alpha={1\over (q-1)!}{1\over p!}A^{i_1\ldots i_{p}}
\alpha_{{i_1}\ldots i_{p} j_1\ldots j_{q-1}} 
dx^{j_1}\wedge\ldots\wedge dx^{j_{q-1}}
\quad,
\eqn\contraction
$$
so that, on forms, $i_Bi_A=i_{A\wedge B}$. 
When $\alpha$ is {\it closed}, eq. \IIi\ provides a definition 
of the SNB through $i_{[A,B]}\alpha$.

>From the definition of the SNB it follows that
$$
[A,B\wedge C]=[A,B]\wedge C+(-1)^{(p-1)q}B\wedge [A,C]\quad,
\eqn\IIii
$$
$$
[A\wedge B,C]=(-1)^p A\wedge [B,C]+ (-1)^{rq}[A,C]\wedge B\quad.
\eqn\IIiia
$$
In particular, for the case of the SNB among the wedge product of two 
{\it vector} fields  
$$
[A\wedge B,X\wedge Y]=
-A\wedge [B,X]\wedge Y 
+B\wedge [A,X]\wedge Y 
-B\wedge [A,Y]\wedge X
+A\wedge [B,Y]\wedge X
\quad,
\eqn\IIiii
$$ 
so that 
$$
[A\wedge B,A\wedge B]=-2 A\wedge B\wedge[A,B]\quad.
\eqn\IIiv
$$
For instance, if $\Lambda$ is given by \IVxiii, we may apply \IIiii\ to find 
that the condition
$[\Lambda,\Lambda]=0$ leads to the Jacobi identity.

\noindent
{\it Remark.\quad}
It is worth mentioning that the SNB is the unique (up to a constant) 
extension of the usual Lie bracket of vector fields (see also Theorem 4.1) 
which makes a $Z_2$-graded Lie algebra 
of the (graded-)commutative algebra of skew-symmetric contravariant tensors:
${\rm degree}([A,B])={\rm degree}(A)+{\rm degree}(B)$.
In it, the adjoint action is 
a graded derivation with respect to the wedge product \refmark{\Koszul} 
(see \IIii).
To make this graded structure explicit, it is convenient to define a new SNB, 
$[\ ,\ ]'$, which differs from the original one $[\ ,\ ]$ by a factor 
$(-1)^{p+1}$ in the r.h.s. of \numerari, \VIIii:
$$
[A,B]':=(-1)^{p+1}[A,B]\quad.
\eqn\nuevosnb
$$
This definition modifies \IIv\ to read
$$
[A,B]'=-(-1)^{(p-1)(q-1)}[B,A]'\equiv -(-1)^{ab}[B,A]'
\quad.
\eqn\IIvb
$$
where $a={\rm degree}(A)=(p-1)$, etc. 
Similarly, \IIvi\ is replaced by
$$
(-1)^{pr+q+1}[[A,B]',C]'+(-1)^{qp+r+1}[[B,C]',A]'+(-1)^{rq+p+1}[[C,A]',B']=0
\quad,
\eqn\IIvib
$$
which in terms of the degrees $(a,b,c)$ of $A,B,C$ adopts the graded JI form 
$$
(-1)^{ac}[[A,B]',C]'+(-1)^{ba}[[B,C]',A]'+(-1)^{cb}[[C,A]',B']=0
\quad.
\eqn\IIvic
$$
The definition \nuevosnb\ is used in \refmark{\Koszul,\GMP,\CIMP,\AP} 
and is more adequate to stress the graded structure of the exterior algebra of 
skew multivector fields; 
for instance, \IIvb\ and \IIvic\ have the same form as in 
supersymmetry (see \eg, \refmark{\CNS}).
In this paper, however, we shall use Def. 4.2 for the SNB as in 
\refmark{\Ni,\Lich,\BFFLS} and others.

\medskip
\noindent{\bf Definition 4.3\quad}
A bivector $\Lambda \in \wedge ^{2}(M)$ is called a
{\it Poisson bivector} 
and defines a PS on $M$ (and a Poisson bracket on 
${\cal F}(M)\times{\cal F}(M)$) if it commutes with itself under the SNB
$$
[\Lambda ,\Lambda ]=0\quad,
\eqn\Ii
$$
(for the case of linear PS 
this is equivalent to the classical Yang--Baxter equation).
Two Poisson bivectors $\Lambda _{1}, \Lambda _{2}$ are called
{\it compatible} if the SNB among themselves is zero, 
$$
[\Lambda _{1}, \Lambda _{2}]=0\quad.
\eqn\Iii
$$
The compatibility condition is equivalent to requiring that any linear 
combination 
$\lambda\Lambda _{1}+\mu \Lambda _{2}$ 
be a Poisson bivector.
 
\chapter{Generalized Poisson structures}

Since \pri\ and \ter\ are automatic for a bivector field, 
the only stringent condition that a 
$\Lambda\equiv\Lambda^{(2)}$ defining a PS must satisfy is the Jacobi 
identity \sec\ or, equivalently, \Ii.
It is then natural to consider generalizations of the standard PS
in terms of $2p$-ary operations determined by
skew-symmetric $2p$-vector fields $\Lambda ^{(2p)}$, the case $p=1$ 
being the standard one. 
Since the SNB vanishes identically if $\Lambda'$ 
is of odd order (eq. \IIv), 
only $[\Lambda',\Lambda']= 0$ for $\Lambda'$ of even order (odd {\it degree}) 
will be non-empty. 

Having this in mind, let us introduce first the GPB.
\medskip
\noindent
{\bf Definition 5.1 \quad} A {\it generalized Poisson bracket}
$\{\cdot ,\cdot ,\ldots ,\cdot ,\cdot \}$ 
on $M$ is a mapping
${\cal F}(M)\times \ld^{2p}\times{\cal F}(M)\to {\cal F}(M)$
assigning a function $\{f_1, f_2,\ldots ,f_{2p}\}$ to every set 
$f_1,\ldots ,f_{2p}\in {\cal F}(M)$
which is linear in
all arguments and satisfies the following conditions:
\medskip
\noindent
a) complete skew-symmetry in $f_j$;
\medskip
\noindent
b) Leibniz rule: $\forall f_i,g,h\in {\cal F}(M),$
$$
\{f_1,f_2,\ldots ,f_{2p-1},gh\} = g\,\{f_1,f_2,\ldots ,f_{2p-1},h\}
+\{f_1,f_2,\ldots ,f_{2p-1},g\}h\quad;
\eqn\fourone
$$
\medskip
\noindent
c) generalized Jacobi identity: $\forall f_i\in {\cal F}(M),$
$$
{1\over (2p-1)!}{1\over (2p)!} 
{\rm Alt}\,\{f_1,f_2,\ldots ,f_{2p-1},\{f_{2p},\ldots ,f_{4p-1}\}\} = 0\quad.
\eqn\fourtwo
$$

Conditions a) and b) imply that our GPB is given by a
skew-symmetric multiderivative \ie, by a completely skew-symmetric
$2p$-vector field $\Lambda ^{(2p)}\in \wedge^{2p}(M)$.
Condition \fourtwo\ (different from the generalization in \refmark{\Na,\Ta})
will be called the {\it generalized
Jacobi identity} (GJI); 
for $p=2$ it contains $35$ terms 
\foot{
Explicitly, the $p=2$ GJI has the form
$$
 \eqalign{&
 \{ f_1,f_2,f_3,\{ f_4,f_5,f_6,f_7\} \}
-\{ f_4,f_2,f_3,\{ f_1,f_5,f_6,f_7\} \}
-\{ f_1,f_4,f_3,\{ f_2,f_5,f_6,f_7\} \}
\cr
&
-\{ f_1,f_2,f_4,\{ f_3,f_5,f_6,f_7\} \}
-\{ f_5,f_2,f_3,\{ f_4,f_1,f_6,f_7\} \}
-\{ f_1,f_5,f_3,\{ f_4,f_2,f_6,f_7\} \}
\cr
&
-\{ f_1,f_2,f_5,\{ f_4,f_3,f_6,f_7\} \}
-\{ f_6,f_2,f_3,\{ f_4,f_5,f_1,f_7\} \}
-\{ f_1,f_6,f_3,\{ f_4,f_5,f_2,f_7\} \}
\cr
&
-\{ f_1,f_2,f_6,\{ f_4,f_5,f_3,f_7\} \}
-\{ f_7,f_2,f_3,\{ f_4,f_5,f_6,f_1\} \}
-\{ f_1,f_7,f_3,\{ f_4,f_5,f_6,f_2\} \}
\cr
&
-\{ f_1,f_2,f_7,\{ f_4,f_5,f_6,f_3\} \}
+\{ f_4,f_5,f_3,\{ f_1,f_2,f_6,f_7\} \}
+\{ f_4,f_2,f_5,\{ f_1,f_3,f_6,f_7\} \}
\cr
&
+\{ f_1,f_4,f_5,\{ f_2,f_3,f_6,f_7\} \}
+\{ f_4,f_6,f_3,\{ f_1,f_5,f_2,f_7\} \}
+\{ f_4,f_2,f_6,\{ f_1,f_5,f_3,f_7\} \}
\cr
&
+\{ f_1,f_4,f_6,\{ f_2,f_5,f_3,f_7\} \}
+\{ f_4,f_7,f_3,\{ f_1,f_5,f_6,f_2\} \}
+\{ f_4,f_2,f_7,\{ f_1,f_5,f_6,f_3\} \}
\cr
&
+\{ f_1,f_4,f_7,\{ f_2,f_5,f_6,f_3\} \}
+\{ f_5,f_6,f_3,\{ f_4,f_1,f_2,f_7\} \}
+\{ f_5,f_2,f_6,\{ f_4,f_1,f_3,f_7\} \}
\cr
&
+\{ f_1,f_5,f_6,\{ f_4,f_2,f_3,f_7\} \}
+\{ f_5,f_7,f_3,\{ f_4,f_1,f_6,f_2\} \}
+\{ f_5,f_2,f_7,\{ f_4,f_1,f_6,f_3\} \}
\cr
&
+\{ f_1,f_5,f_7,\{ f_4,f_2,f_6,f_3\} \}
+\{ f_6,f_7,f_3,\{ f_4,f_5,f_1,f_2\} \}
+\{ f_6,f_2,f_7,\{ f_4,f_5,f_1,f_3\} \}
\cr
&
+\{ f_1,f_6,f_7,\{ f_4,f_5,f_2,f_3\} \}
-\{ f_4,f_5,f_6,\{ f_1,f_2,f_3,f_7\} \}
-\{ f_4,f_5,f_7,\{ f_1,f_2,f_6,f_3\} \}
\cr
&
-\{ f_4,f_6,f_7,\{ f_1,f_5,f_2,f_3\} \}
-\{ f_5,f_6,f_7,\{ f_4,f_1,f_2,f_3\} \}=0
\quad.
\cr}
$$
}
($C^{2p-1}_{4p-1}$ in the general case). 
It may be rewritten as $[\Lambda ^{(2p)}, \Lambda ^{(2p)}]
=0$ which, due to \IIv, is not identically zero and gives a non-trivial 
condition;
$\Lambda^{(2p)}$ 
defines a GPB.
We shall see in Sec. 8 that in the linear case our generalized PS are 
automatically obtained from {\it constant} skew-symmetric tensors of order 
$2p+1$.
Clearly, the above relations reproduce the ordinary case \pri--\sec\ for
$p=1$. The compatibility condition in Def. 4.3
may be now extended in the following sense:
two GPS $\Lambda_1 ^{(2p)}$ and ${\Lambda}_2^{(2q)}$ on $M$
are called {\it compatible} if they `commute' \ie, $[\Lambda_1 ^{(2p)},
{\Lambda}_2 ^{(2q)}]=0$ 
(of course, if $p\neq q$ the sum of $\Lambda_1 ^{(2p)}$ and 
${\Lambda}_2 ^{(2q)}$ is not defined).

In local coordinates the GPB has the form
$$
\{f_1(x), f_2(x),\ldots ,f_{2p}(x)\}=\omega _{j_1j_2\ldots j_{2p}}(x)
\partial^{j_1}f_1\,\partial^{j_2}f_2\,\ldots \,\partial^{j_{2p}}f_{2p}\;\quad.
\eqn\fourthree
$$
where 
$\omega_{j_1j_2\ldots j_{2p}}$ 
are the coordinates of a completely skew-symmetric tensor which,
as a result of \fourtwo, satisfies
$$
{\rm Alt}\,(\omega _{j_1j_2\ldots j_{2p-1}k}\,\partial ^k\,
\omega_{j_{2p}\ldots j_{4p-1}}) = 0\quad.
\eqn\fourfour
$$

\medskip
\noindent
{\bf Definition 5.2\quad}
A skew-symmetric $2p$-vector field 
$\Lambda ^{(2p)}\in \wedge^{(2p)}(M)$, locally written as 
$$
\Lambda ^{(2p)} = {1\over {(2p)!}}\,\omega _{j_1\ldots j_{2p}}\,\partial^{j_1}
\land \ldots \land \partial ^{j_{2p}}\quad,
\eqn\fourfive
$$
defines a {\it generalized Poisson structure} {\it iff}
$[\Lambda ^{(2p)},\Lambda ^{(2p)}]=0$, which reproduces eq. \fourfour).

\chapter{Generalized dynamics}

Let us now introduce a dynamical system associated with the above generalized 
Poisson structure. 
Namely, let us fix a set of $(2p-1)$ `Hamiltonian' functions 
$H_1,H_2,\ldots , H_{2p-1}$.
The time evolution of $x_j$, $f\in{\cal F}(M)$ is defined by
$$
\dot x_j=\{H_1,\ldots ,H_{2p-1}, x_j\}\quad,\quad
\dot f=\{H_1,\ldots ,H_{2p-1}, f\}\quad.
\eqn\timeevolution
$$
\medskip
\noindent
{\bf Definition 6.1\quad}
The {\it Hamiltonian vector field} associated with the $(2p-1)$ Hamiltonians 
$H_1,\ldots,H_{2p-1}$ is defined by 
$X_{H_1,\ldots,H_{2p-1}}=i_{dH_1\wedge\ldots\wedge dH_{2p-1}}\Lambda$.
Thus,
$$
\eqalign{
(i_{dH_1\wedge\ldots\wedge dH_{2p-1}}\Lambda)_{j}=
&
{1\over (2p-1)!}\Lambda(dH_1\wedge\ldots\wedge dH_{2p-1},dx_j)
\cr
=
&
\Lambda(dH_1,\ldots, dH_{2p-1},dx_j)\quad;
\cr
X_{H_1,\ldots,H_{2p-1}}=
&
\Lambda_{i_1\ldots i_{2p-1} j}
\partial^{i_1}H_1\ldots\partial^{i_{2p-1}}H_{2p-1}\partial^j\quad.
\cr}
\eqn\hamvecf
$$

\medskip
\noindent
{\bf Definition 6.2\quad}
The {\it generalized Hamiltonian system} is defined by the equation
$$
\dot x_j=X_j=(X_{H_1,\ldots,H_{2p-1}})_j=
\Lambda_{i_1\ldots i_{2p-1} j}
\partial^{i_1}H_1\ldots\partial^{i_{2p-1}}H_{2p-1}\quad,
\eqn\generalsystem
$$
Then, $\dot f=X_{H_1,\ldots H_{2p-1}}.f\ 
(=\dot x_j{\partial f\over\partial x_j})$
is given by \timeevolution.

\medskip
\noindent
{\bf Definition 6.3\quad}
A function $f\in{\cal F}(M)$ is a {\it constant of the motion} 
if \timeevolution\ is zero.

Due to the skew-symmetry of the GPB, the Hamiltonian
functions 
$H_1,\ldots,\allowbreak H_{2p-1}$ 
are all constants of the motion but the system may have additional ones 
$h_{2p},\ldots ,\allowbreak h_k\,;\,k\geq 2p$. 

\medskip
\noindent
{\bf Definition 6.4\quad}
A set of functions $(f_1,\ldots ,f_k)\,,k\geq 2p$ 
is in {\it involution} if the GPB vanishes for any subset of $2p$ functions.

Let us note also the following generalization of the Poisson theorem
\refmark{\Po}:

\medskip
\noindent
{\bf Theorem 6.1\quad}
Let $f_1,\ldots ,f_{q}\,,\, q\geq 2p$ be such that the
set of functions 
$(H_{1},\ldots ,\allowbreak H_{2p-1},
f_{i_{1}},\ldots ,f_{i_{2p-1}})$ is in involution 
(this implies, in particular, that the $f_i\,,\, i=1,\ldots,q$ 
are constants of motion).
Then, the
quantities $\{f_{i_1},\ldots ,f_{i_{2p}}\}$ are also constants of motion.

\medskip
\noindent
{\bf Definition 6.5\quad}
A set of $k$ functions 
$c_1(x),\ldots,c_k(x)\,\, (1\leq k\leq 2p-1)$ 
will be called a set of $k$ {\it Casimir functions} if 
$\{g_1, g_2,\ldots ,\allowbreak g_{2p-k},c_1,\ldots, c_k\}=0$ 
for any set of functions 
$(g_1, g_2,\ldots ,g_{2p-k})$.

If one of the Hamiltonians 
$(H_1,\ldots ,H_{2p-1})$ is a Casimir function, then 
the generalized dynamics defined by \timeevolution\ is trivial.
Also, if the set of Hamiltonians contains a Casimir subset, the generalized 
dynamics will also be trivial (note that if $H_1$ and $H_2$ constitute 
{\it each} a Casimir subset, the two Hamiltonians $(H_1,H_2)$ will also 
constitute another, but the reciprocal situation may not be true).

Each Casimir $k$-subset $(c_1(x),\ldots,c_k(x))$ determines invariant 
submanifolds of $M$ through the conditions $c_i(x)=c_i\ (i=1,\ldots, k)$.
The maximal $K$-subset determines an invariant submanifold of $M$ of
minimal dimension, dim$M-K$, which we may call phase space.
Using now the notion of support of an $m$-skew multivector
\refmark{\AG} as the subspace of the space of vector fields generated by the
contraction of the multivector with an arbitrary $(m-1)$-form, 
we make the following

\noindent
{\it Conjecture:\quad} 
The tangent space to the phase space at a point $x\in M$ is the support of 
$\Lambda(x)$ at that point.

\medskip
\noindent
{\it Remark.}\quad
It is well known that the standard Jacobi identity among $f_1\,,\,f_2$ and $H$ 
is equivalent to 
${d\over dt}\{f_1,f_2\}=\{\dot f_1,f_2\}+\{f_1,\dot f_2\}$; 
thus, $d/dt$ is a derivation of the PB. 
The `fundamental identity' for Nambu mechanics \refmark{\Na} and its further 
extensions \refmark{\Ta} also correspond to the existence of a vector field 
$D_{H_1\ldots H_{k-1}}$ which is a derivation of the Nambu bracket.
In contrast, the vector field \generalsystem\ above is not a derivation
of our GPB.
It should be noticed, however, that having an evolution vector field which is a 
derivation of a PB is an independent 
assumption of the associated dynamics and not a necessary one.
Nevertheless, the following theorem holds:

\medskip
\noindent
{\bf Theorem 6.2}\quad
Let $H_1,\ldots, H_{2p-1}$ be the `Hamiltonians' governing the time evolution 
by \timeevolution\ and let $f_1,\ldots, f_{2p}$ a set of $2p$ functions such that any 
subset $(f_{i_1},f_{i_2},\allowbreak
\ldots,f_{i_{2p-1}},H_{j_1},\ldots,H_{j_{2p-2}})$ is 
in involution.
Then,
$$
{d\over dt}\{f_1,\ldots,f_{2p}\}=\{\dot f_1,f_2,\ldots,f_{2p}\}+\ldots+
\{f_1,\ldots,f_{2p-1},\dot f_{2p}\}
\eqn\theorem
$$
\medskip
\noindent
{\it Proof.}\quad
It suffices to check that \timeevolution\ in \theorem\ leads to an identity on 
account of the generalized Jacobi identity \fourtwo.
For the case $p=2$, for instance, the condition of the theorem (see previous 
footnote) states that any GPB involving two Hamiltonians and two functions or 
one Hamiltonian and three functions is zero.

\medskip
\noindent
{\bf Example}\quad
It is well known that Euler's equations describing the free motion of a rigid 
body around a fixed point are Hamiltonian, $\dot x_i=\{H,x_i\}$, where
$H\propto a_1 x_1^2+ a_2 x_2^2+a_3 x_3^2$ (where $a_i$ are the principal 
moments of the body) and the (linear) PS is defined by
$\{x_i,x_j\}=\epsilon_{ij}^{~~k}x_k$ so that 
$\dot x_j\propto \epsilon_{ij}^{~~k}\partial ^i H x_k$, $i,j,k=1,2,3$.
The extension of this situation to the motion in a $(2p+1)$-dimensional space 
provides an example of our GPS.
Let the evolution equations be given in terms of $(2p-1)$ Hamiltonians 
$H_1,\ldots,H_{2p-1}$ (the above case corresponds to $p=1$) by
$$
\dot x_j=\epsilon_{i_1\ldots i_{2p-1} j k}
\partial^{i_1}H_1\ldots \partial^{i_{2p-1}}H_{2p-1} x^k\quad,\quad
i,j,k=1,...,2p+1\quad.
\eqn\examplea
$$
These equations have the Hamiltonian form \timeevolution\ if the PS is the 
linear one defined by
\foot{
It is worth mentioning that the completely antisymmetric tensor of order 
$(n+1)$ in a $(n+1)$-dimensional vector space gives 
rise \refmark{\ChaTak} to a Nambu tensor 
$\epsilon_{i_1\ldots i_n i_{n+1}} x^{n+1}$ 
of order $n$ 
(\ie, to a tensor satisfying the 
`fundamental identity' of Nambu mechanics \refmark{\Ta}),
so that $\omega_{i_1\ldots i_{2p}}(x)$ above is also a Nambu 
tensor.
}
$$
\{x_{i_1},x_{i_2},\ldots,x_{i_{2p}}\}=\epsilon_{i_1\ldots i_{2p} k}x^k\equiv
\omega_{i_1\ldots i_{2p}}(x)\quad.
\eqn\exampleb
$$
Due to the form of $\omega_{i_1\ldots i_{2p}}(x)$, it is clear that the GJI 
\fourfour\ 
is trivially fulfilled, since it will always involve the antisymmetrization of 
repeated indices.
Thus, \exampleb\ defines a linear GPS reproducing \examplea;
clearly the $(2p-1)$ Hamiltonians are constants of the motion.
As in the 3-dimensional analogue, the function determining a $S^{2p}$-sphere 
$c_{2p}=x_1^2+\ldots+x_{2p+1}^2$ is a Casimir function (and a constant of 
motion).
Indeed (Def. 6.5),
$$
\{f_1,\ldots,f_{2p-1},x_1^2+\ldots+x_{2p+1}^2\}=2\omega_{i_1\ldots i_{2p}}
\partial^{i_1}f_{1}\ldots\partial^{i_{2p-1}}f_{2p-1} x^{i_{2p}}
\eqn\examplec
$$
which is zero for all $f$'s on account of \exampleb.
The trajectory is thus the intersection of the surfaces $H_l=$const. 
$(l=1,\ldots,2p-1)$ and $c_{2p}=$const.

Eqs. \examplea\ are not quadratic in $x^i$ in general and so they do not 
coincide with the standard Euler equations for the rotation of a higher 
dimensional rigid body.
They become quadratic when $H_1,\ldots,H_{2p-2}$ are linear and $H_{2p-1}$ is 
a quadratic function of the coordinates, but in this case they reduce to the 
standard Euler equations in the three-dimensional space determined by the 
intersection of the $H_i=$const. ($i=1,\ldots,2p-2$) hyperplanes.

\chapter{Generalized Poisson structures and differential forms}

Let us now rewrite some of the previous expressions in terms of
differential forms.
First we associate 
$(n-k)$-forms $\alpha$ with $k$-skew-symmetric contravariant  
tensor fields $\Lambda$ on an $n$-dimensional orientable 
manifold $M$ by setting
$$
\alpha _\Lambda = i_\Lambda \mu\quad,
\eqn\aIVi
$$
where $\mu$ stands for a volume form on $M$ (hence, $\alpha_\Lambda$ depends 
on the choice of the volume form $\mu$). 
The mapping $\Psi:\Lambda\mapsto\alpha_\Lambda$ 
yields an isomorphism between $k$-skew multivectors and $(n-k)$--forms.
For a $\Lambda$ given by the exterior product of $k$ vector fields 
$\Lambda =X_1\wedge ...\wedge X_k$ (see \contraction),
$$
(i_\Lambda \mu )(Y_1,...,Y_{n-k})=\mu (X_1,...,X_k,
Y_1,...,Y_{n-k})\quad.
\eqn\aIVii
$$
Locally, if for example $\Lambda ={1\over 2}\omega^{ij}\partial _i\wedge 
\partial _j$ and
$\mu =dx^1\wedge ...\wedge dx^n$, eq. \aIVii\ gives
$$
\alpha_\Lambda =\sum_{i<j}(-1)^{i+j+1}\omega^{ij}dx^1\wedge ...
{\widehat {dx^i}}...{\widehat {dx^j}}...\wedge dx^n\quad,
\eqn\numerariv
$$
where ${\widehat {dx^i}}$ stands for the omission of $dx^i$.

For vector fields $X,Y$, 
$$
i_{[X,Y]}=i_Xdi_Y-i_Ydi_X+i_Xi_Yd-di_{X\wedge Y}\quad;
\eqn\numerarv
$$
similarly, for two bivector fields $\Lambda _1,\ \Lambda _2$, 
$$
i_{[\Lambda _1,\Lambda _2]}=i_{\Lambda _1}di_{\Lambda _2}+
i_{\Lambda _2}di_{\Lambda _1}
-i_{\Lambda _1}i_{\Lambda _2}d-
di_{\Lambda _1\wedge \Lambda _2}\quad.
\eqn\numerarvi
$$
In general, for any two skew-symmetric multivectors $A,B$ of order $p,q$ 
acting on forms we have
\foot{
Eq. \acting\ is to be compared with the standard formula for vector fields
$i_{[X,Y]}=[L_X,i_Y]=i_Xdi_Y-i_Ydi_X+i_Xi_Yd+di_Xi_Y$ to which it reduces for 
$p=1=q$ (on forms, $i_Ai_B=(-1)^{pq}i_Bi_A$ and 
$i_{A\wedge B}=(-1)^{pq}i_{B\wedge A}=(-1)^{pq}i_Ai_B$).
One could introduce a Lie `derivative' $L_A$ with respect $A\in\wedge^p(M)$
and rewrite \acting\ in a form similar to the vector field case, namely
$i_{[A,B]}=\lbr L_A,i_B \rbr$, where $L_A:=i_Ad+(-1)^{p+1}di_A\ 
(L_A:\wedge^n\to\wedge^{n-p+1}$ and thus it is a derivative only if $A$ is 
a vector field) and the bracket $\lbr\ ,\ \rbr$ is defined
by $\lbr L_A, i_B\rbr := (-1)^{q(p+1)}L_Ai_B-i_BL_A$.
}
$$
i_{[A,B]}=(-1)^{pq+q}i_Adi_B+(-1)^{p}i_Bdi_A
-(-1)^{pq}i_Ai_Bd-(-1)^{p+q}di_{A\wedge B}\quad,
\eqn\acting
$$
from which we find that \numerarvi\ remains valid for any two skew-symmetric 
multivectors $\Lambda_1$ and $\Lambda_2$ of {\it even} order
(on a $(p+q-1)$-form $\alpha$, \acting\ reduces to \IIi\ since $A\wedge B\in 
\wedge^{p+q}$).
Eq. \numerarvi\ now leads to the following theorem which 
generalizes that in \refmark{\GMP}
to the arbitrary even order case:
\medskip
\noindent{\bf Theorem 7.1\quad}
A $\Lambda$ defines a GPS if and only if 
$$
2i_{\Lambda}d\alpha _{\Lambda} = d\alpha _{\Lambda \wedge \Lambda}
\quad.
\eqn\IIIi
$$
Two GPS $\Lambda _1,\ \Lambda _2$ are compatible if 
and only if 
$$
d\alpha_
{\Lambda _1\wedge \Lambda _2} 
=i_{\Lambda _1}d\alpha_{\Lambda _2} + i_
{\Lambda _2}d\alpha_{\Lambda _1}\quad.
\eqn\numerarvia
$$

The isomorphism defined by $\Psi$ suggests to compose it with 
the differential operators $d,\ L_X,\ i_X$ 
available on forms, 
so that the properties of the Schouten bracket can be stated
in terms of differential forms. We have, as it is well known,
$$
L_X\mu =div(X)\mu =di_X\mu\quad,
\eqn\IIIii
$$
$$
d(i_{X\wedge Y}\mu )=i_{[Y,X]}\mu +i_Xdi_Y\mu-i_Ydi_X\mu\quad.
\eqn\IIIiii
$$
Thus, defining 
$D=\Psi ^{-1}\circ d\circ \Psi ,$ we get for vector fields $X,Y$
$$
D(X)=div(X)\quad,
\eqn\IIIiv
$$
$$
D(X\wedge Y)=-div(X)Y+Xdiv(Y)-[X,Y]\quad.
\eqn\IIIv
$$
For contravariant, skew-symmetric tensor fields $\Lambda _1,\Lambda _2$ 
of arbitrary even order we obtain 
$$
D(\Lambda _1\wedge \Lambda _2)=D(\Lambda _1)\wedge \Lambda _2
+\Lambda _1\wedge D(\Lambda _2)-[\Lambda _1,\Lambda _2]\quad,
\eqn\numerarxx
$$
and in the general case we have
$$
D(A\wedge B)=(-1)^q D(A)\wedge B +A\wedge D(B)-
(-1)^{p+q}[A,B]\quad.
\eqn\Dacting
$$
Hence we conclude that if $\Lambda$ is of arbitrary even order and defines 
a GPS,
$$
D(\Lambda \wedge \Lambda)=2\Lambda \wedge D(\Lambda)\quad.
\eqn\IIIvi
$$
We may call a GPS $\Lambda$ {\it closed} if 
$D(\Lambda )=0$ (which implies $D(\Lambda \wedge \Lambda)=0$). This is clearly 
equivalent to the fact that the form $\alpha_\Lambda$ (and hence 
$\alpha_{\Lambda 
\wedge \Lambda}$) is closed. 
As mentioned, this definition depends on $\mu$ so that if 
$\mu$ is replaced by 
$f\mu$, $\Lambda$ may be no longer closed.

\chapter{The Schouten-Nijenhuis bracket, GPS and cohomology}

Let $\Lambda^{(2p)}$ be a $(2p)$-skew-symmetric multivector defining a GPS as 
in Def. 5.2.
Using \IIvi\ it follows that the mapping 
$\delta_{\Lambda^{(2p)}}:B\mapsto[\Lambda,B]\,,\,
\delta_{\Lambda^{(2p)}}:\wedge^q(M)\to\wedge^{2p+q-1}(M)$ is nilpotent since 
$[\Lambda,[\Lambda,B]]=0$.
We then have the following

\medskip
\noindent
{\bf Theorem 8.1\quad}
Let a GPS be defined by $\Lambda^{(2p)}$.
The mapping $\delta_{\Lambda^{(2p)}}:B\mapsto [\Lambda,B]$ is nilpotent,
$\delta^2_{\Lambda^{(2p)}}=0$. 
The operator $\delta_{\Lambda^{(2p)}}$ satisfies (see \IIii, \IIvi)
$$
\delta_{\Lambda^{(2p)}}(B\wedge C)=
(\delta_{\Lambda^{(2p)}} B)\wedge C+ (-1)^{q}B\wedge(\delta_{\Lambda^{(2p)}}C)
\quad
\eqn\cohomology
$$
$$
\delta_{\Lambda^{(2p)}}[B,C]=-[\delta_{\Lambda^{(2p)}}B,C]-(-1)^{q}
[B,\delta_{\Lambda^{(2p)}}C]\quad.
\eqn\mascohomology
$$
As a result, $\delta_{\Lambda^{(2p)}}$ defines an odd degree cohomology 
operator; the resulting cohomology will be called {\it generalized Poisson 
cohomology}. 
In particular, for $p=1$, 
$\delta_{\Lambda^{(2)}}:\wedge^q(M)\to\wedge^{q+1}(M)$ 
defines the standard Poisson cohomology \refmark{\Lich}; see also 
\refmark{\Koszul}.

Let us now turn to linear GPS.
Let $\g$ be the Lie algebra 
of a simple compact group $G$. In this case the de Rham 
cohomology ring on the group manifold $G$ is the same as the Lie algebra 
cohomology ring $H^*_0(\g,\R)$ for the trivial action. In its
Chevalley-Eilenberg version \refmark{\CE} 
the Lie algebra cocycles are represented by 
bi-invariant (\ie, left and right invariant and hence closed) forms on $G$
(see also, \eg, \refmark{\AI}).
The linear standard PS defined by \IVvii\ is associated 
(see \refmark{\APPBJPA}) 
with a non-trivial three-cocycle on
$\g$ and $[\Lambda^{(2)},\Lambda^{(2)}]=0$ (eq. \VIa) is precisely
the cocycle condition. 
This indicates that the linear generalized Poisson 
structures on $\g^*$ may be found by looking for higher order Lie algebra 
cocycles.
Let us now show that each of them provides a GPS.

The cohomology ring of any simple Lie algebra of rank $l$ is a free ring 
generated by $l$ (primitive) forms on $G$ of odd degree $(2m-1)$. 
These forms are associated 
with the $l$ primitive symmetric invariant tensors $k_{i_1\ldots i_m}$ 
of order $m$ which may be defined on 
$\g$ and of which the Killing tensor $k_{i_1i_2}$ is just the 
first example (and thus $H^3_0(\g,\R)\neq 0$ for any simple Lie 
algebra).
As a result, it is possible to associate a $(2m-2)$ 
skew-symmetric contravariant primitive 
tensor field linear in $x_j$ 
to each symmetric invariant polynomial $k_{i_1\ldots i_m}$ 
of order $m$.
The case $m=2$ leads to the $\Lambda^{(2)}$ of \IVvii, \IVxiii.
For the $A_l$ series $(su(l+1))$, for instance, these forms have order 
$3,5,\ldots,(2l+1)$; other orders (but always including $3$)
appear for the different simple algebras 
(see, \eg, \refmark{\AI}).
Let $\{e_i\}$ be a basis of $\g$. The bi-invariance condition
$$
\sum^{q}_{s=1}
\omega(e_{i_1},\ldots,[e_l,e_{i_s}],\ldots,e_{i_{q}})=0
\eqn\VIiii
$$
reads, in terms of the coordinates 
$\omega_{i_1\ldots i_{q}}=\omega(e_{i_1},\ldots,e_{i_q})$ 
of the skew-symmetric tensor $\omega$ on $\g$ (or, equivalently, 
LI $q$-form $\omega$ on $G$),
$$\eqalign{
C^\alpha_{li_1}\omega_{\alpha i_2\ldots i_{q}}+
C^\alpha_{li_2}\omega_{i_1\alpha i_3\ldots i_{q}}+
&\ldots +
C^\alpha_{li_{q}}\omega_{i_1\ldots i_{q-1}\alpha}=0
\quad,\cr
\sum^{q}_{s=1}
C_{\nu i_s}^{\rho}&\omega_{i_1\ldots\hat{i_s}\rho\ldots i_{q}}=0\quad.\cr}
\eqn\VIiv
$$
The bi--invariance 
condition may also be expressed as
$$
\epsilon^{j_1\ldots j_{q}}_{i_1\ldots i_{q}}
C^{\rho}_{\nu j_1}\omega_{\rho j_2\ldots j_q}=0\quad,
\eqn\VIxi
$$
on account of the skew-symmetry of $\omega$.
Using the Killing metric this leads to
$$
\epsilon^{j_1\ldots j_{q}}_{i_1\ldots i_{q}}
C^{\nu}_{j_1 \rho}\omega^{\rho}_{~j_2\ldots j_q}=0\quad.
\eqn\VIxii
$$

Let $\omega$ be a Lie algebra $q$-cochain (\ie\ a skew--symmetric $q$-tensor
on $\g$ or LI $q$-form on $G$). The coboundary operator 
for the Lie algebra cohomology is given by

\medskip
\noindent
\def\sums{\sum^{q+1}_{\scriptstyle s,t=1 \atop \scriptstyle s<t}}
{\bf Definition 8.1a}\quad ({\it Coboundary operator})
$$
(s\omega)(e_{i_1},\ldots,e_{i_{q+1}}):=
\sums (-1)^{s+t}\omega([e_{i_s},e_{i_t}],e_{i_1},\ldots,\hat{e_{i_s}},\ldots, 
\hat{e_{i_t}},\ldots,e_{i_{q+1}})\;,\; e_i\in\g
\eqn\VIviii\;
$$
Thus, in coordinates,
$$
\eqalign{(s\omega)_{i_1\ldots i_{q+1}}=&
\sums(-1)^{s+t}C^\rho_{i_s i_t}\omega_{\rho i_1\ldots\hat{i_s}\ldots\hat{i_t}
\ldots i_{q+1}}\cr
=&
{1\over 2}
\sums(-1)^{s+t}\epsilon^{j_1j_2}_{i_s i_t} C^\rho_{j_1 j_2}
\omega_{\rho i_1\ldots\hat{i_s}\ldots\hat{i_t}
\ldots i_{q+1}}\cr
=&
{1\over 2}\sums(-1)^{s+t}
\epsilon^{j_1j_2}_{i_s i_t} 
C^\rho_{j_1 j_2}
{1\over (q-1)!}
\epsilon^{j_3\ldots j_{q+1}}_
{\rho i_1\ldots\hat{i_s}\ldots\hat{i_t}\ldots i_{q+1}}
\omega_{\rho j_3\ldots j_{q+1}}\cr
=&
-{1\over 2}{1\over (q-1)!}C^\rho_{j_1 j_2}\omega_{\rho j_3\ldots j_{q+1}}
\sums (-1)^{s+t+1}
\epsilon^{j_1j_2}_{i_s i_t} 
\epsilon^{j_3\ldots j_{q+1}}_
{\rho i_1\ldots\hat{i_s}\ldots\hat{i_t}\ldots i_{q+1}}\quad.
\cr}
\eqn\VIix
$$
This provides the equivalent 
\medskip
\noindent
{\bf Definition 8.1b\quad}
The action of the coboundary operator on a $q$-cochain $\omega$ is given by
$$
(s\omega)_{i_1\ldots i_{q+1}}=
-{1\over 2}{1\over (q-1)!}
\epsilon^{j_1\ldots j_{q+1}}_{i_1\ldots i_{q+1}}
C^\rho_{j_1 j_2}\omega_{\rho j_3\ldots j_{q+1}}
\quad;\quad s\omega=0\ {\rm for}\ q>r={\rm dim}\g\quad.
\eqn\VIx
$$

As is well known, the invariance condition \VIiii\ determines a Lie algebra 
cocycle since, for each fixed $j_1$, the antisymmetric sum over 
$j_2,\ldots,j_n$ is zero on account of \VIiii.
The Poisson structure \IVxii\ 
is associated to the structure constants and hence to a three-cocycle.
In order to obtain more general structures, we need the expression of 
the $(2m-1)$-cocycle associated with an order $m$ symmetric tensor on \g.
This is done in two steps, the first of which is provided by  the following
\medskip
\noindent
{\bf Lemma 8.1\quad}
Let
$k_{i_1\ldots i_m}$ be
an invariant symmetric polynomial on $\g$ and 
$$
\tilde\omega_{\rho j_2\ldots j_{2m-2}\sigma}:=
k_{i_1\ldots i_{m-1}\sigma}
C^{i_1}_{\rho j_2}\ldots C^{i_{m-1}}_{j_{2m-3}j_{2m-2}}\quad.
\eqn\VIxiii$$
Then, the odd order $(2m-1)$-tensor
$$
\omega_{\rho l_2\ldots l_{2m-2} \sigma}:=
\epsilon^{j_2\ldots j_{2m-2}}_{l_2\ldots l_{2m-2}}
\tilde\omega_{\rho j_2\ldots j_{2m-2} \sigma}
\eqn\VIxiv
$$
is a fully skew-symmetric tensor
\foot
{The origin of \VIxiv\ follows from the fact that given a symmetric 
invariant polynomial $k_{i_1\ldots i_m}$ on $\g$, the associated 
skew-symmetric multilinear tensor $\omega_{i_1\ldots i_{2m-1}}$ 
is
$$
\omega(e_{i_1},\ldots, e_{i_{2m-1}})=
\sum_{s\in S_{(2m-1)}}{\pi(s)}
k([e_{s(i_1)},e_{s(i_2)}],
[e_{s(i_3)},e_{s(i_4)}],\ldots,
[e_{s(i_{2m-3})},e_{s(i_{2m-2})}],e_{s(i_{2m-1})})
$$
where $\pi(s)$ is the parity sign of the permutation $s\in S_{(2m-1)}$.}.

\noindent
{\it Proof}.\quad 
For $m=2$, the skew-symmetry of 
$\omega_{\rho j_2 \sigma}=k_{i_1\sigma}C^{i_1}_{\rho j_2}$ 
is obvious. In general,
$$\eqalign{&
\epsilon^{j_2\ldots j_{2m-2}}_{l_2\ldots l_{2m-2}}
\tilde\omega_{\rho j_2\ldots j_{2m-2}\sigma}=
\epsilon^{j_2\ldots j_{2m-2}}_{l_2\ldots l_{2m-2}}
k_{i_1\ldots i_{m-1}\sigma}
C^{i_1}_{\rho j_2}\ldots C^{i_{m-1}}_{j_{2m-3}j_{2m-2}}\cr
&=
\epsilon^{j_2\ldots j_{2m-2}}_{l_2\ldots l_{2m-2}}
\left[
\sum_{s=2}^{m-1}k_{\rho i_2\ldots\hat{i_s}i_1\ldots i_{m-1}\sigma}
C^{i_1}_{j_2i_s} + k_{\rho i_2\ldots i_{m-1}i_1}C^{i_1}_{j_2\sigma}
\right]
C^{i_2}_{j_3j_4}\ldots C^{i_{m-1}}_{j_{2m-3}j_{2m-2}}\cr
&=
\epsilon^{j_2\ldots j_{2m-2}}_{l_2\ldots l_{2m-2}}
k_{\rho i_2\ldots i_{m-1}i_1}C^{i_1}_{j_2\sigma}
C^{i_2}_{j_3j_4}\ldots C^{i_{m-1}}_{j_{2m-3}j_{2m-2}}\cr
&=
-\epsilon^{j_2\ldots j_{2m-2}}_{l_2\ldots l_{2m-2}}
k_{i_1 i_2\ldots i_{m-1}\rho}C^{i_1}_{\sigma j_2}\ldots
C^{i_{m-1}}_{j_{2m-3}j_{2m-2}}=
-\epsilon^{j_2\ldots j_{2m-2}}_{l_2\ldots l_{2m-2}}
\tilde\omega_{\sigma j_2\ldots j_{2m-2}\rho}\quad,\cr}
$$
where the invariance of the symmetric tensor $k$ has been used in the second 
equality, the Jacobi identity in the third and the symmetry of $k$ in the 
fourth. Since $\omega_{\rho j_2\ldots j_{2m-2} \sigma}$ is skew-symmetric in 
$(\rho, \sigma)$ it follows that $\omega_{i_1\ldots i_{2m-1}}$ 
is a fully skew-symmetric
tensor {\it q.e.d.}
We may then state the following
\medskip
\noindent
{\bf Theorem 8.2\quad}
The skew-symmetric tensor $\omega_{i_1\ldots i_{2m-1}}$ on $\g$ (or LI 
$(2m-1)$--form on $G$) of \VIxiv\ is a $(2m-1)$-cocycle for the Lie algebra 
cohomology. 

\noindent
{\it Proof}.\quad  Applying eq. \VIx\ to 
\VIxiv\ and using \VIxiii\ it follows that
$$
\eqalign{
(s\omega)_{i_i\ldots i_{2m}}=&
-{1\over 2(2m-2)!}\epsilon^{j_1\ldots j_{2m}}_{i_1\ldots i_{2m}}
C^\rho_{j_1j_2}
\epsilon^{s_3\ldots s_{2m-1}}_{j_3\ldots j_{2m-1}}
k_{l_1 l_2\ldots l_{m-1} j_{2m}}C^{l_1}_{\rho s_3}\ldots
C^{l_{m-1}}_{s_{2m-2}s_{2m-1}}\cr
=&
-{(2m-3)!\over 2(2m-2)!}
\epsilon^{j_1 j_2 s_3\ldots s_{2m-1} j_{2m}}_{i_1\ldots i_{2m}}
C^\rho_{j_1j_2}
C^{l_1}_{\rho s_3}\ldots
C^{l_{m-1}}_{s_{2m-2}s_{2m-1}}
k_{l_1 l_2\ldots l_{m-1} j_{2m}}
\cr
=&0
\cr}
$$
by the Jacobi identity \VIa\ in the two first structure constants 
(indices $j_1,j_2,s_3$), {\it q.e.d.}
\medskip
\noindent
{\bf Lemma 8.2\quad}
Let $\omega_{i_1\ldots i_q}$, $q$ odd, be an skew-symmetric tensor 
associated with an invariant symmetric polynomial as above. Then
$$
\epsilon^{j_1\ldots j_q}_{i_1\ldots i_q}C^\rho_{j_1 j_2}
\omega_{\rho j_3\ldots j_q \nu}=0\quad.\eqn\VIxv
$$
{\it Proof}.\quad By \VIxiv\ and \VIxiii
$$\eqalign{&
{1\over(q-2)!}
\epsilon^{j_1\ldots j_q}_{i_1\ldots i_q}C^\rho_{j_1 j_2}
\epsilon^{l_3\ldots l_q}_{j_3\ldots j_q}
\tilde\omega_{\rho l_3\ldots l_q \nu}
=
\epsilon^{j_1 j_2 l_3\ldots l_q}_{i_1\ldots i_q}
C^\rho_{j_1 j_2} 
C^{s_1}_{\rho l_3}\ldots C^{s_p}_{l_{q-1} l_q}k_{s_1\ldots s_p\nu}\quad,\cr}
$$
where $p=(q-1)/2$, which is zero on account of the Jacobi identity for 
$j_1,j_2,l_3$, {\it q.e.d.}

The possible cocycles on the different simple Lie algebras are determined by 
the symmetric invariant polynomials that may be defined on them, which in turn 
are in one-to-one correspondence with the non-trivial de Rham cocycles, which 
exist on the corresponding compact group manifolds.
Using the above constructions, we may now introduce higher order ($>2$) 
contravariant skew-symmetric tensors which have zero SNB among themselves.

Let us now apply the results of Sec. 8 to compute the SNB of two contravariant 
skew-symmetric tensors $\Omega$ and $\Omega'$ obtained  
from Lie algebra cocycles,
$$
\Omega:={1\over p!}{\omega_{i_i\ldots i_p}}^\alpha x_\alpha 
\partial^{i_1}\wedge\ldots\wedge \partial^{i_p}\quad,\quad
\Omega':={1\over q!}{\omega'_{j_i\ldots j_q}}^\alpha x_\alpha 
\partial^{j_1}\wedge\ldots\wedge \partial^{j_q}\quad,
\eqn\VIIi
$$
where $x^\alpha\in\g^*$.
Using the Killing metric to raise and lower indices, eq. \VIIii\ gives
$$
\eqalign{
[\Omega,\Omega']_{i_1\ldots i_{p+q-1}}=&
\left\{
{1\over (p-1)!q!}\epsilon^{j_1\ldots j_{p+q-1}}_{i_1\ldots i_{p+q-1}}
\omega_{\nu j_1\ldots j_{p-1}\alpha}{\omega'_{j_p\ldots j_{p+q-1}}}^\nu
\right.
\cr
&
\left.
{(-1)^p\over p!(q-1)!}
\epsilon^{j_1\ldots j_{p+q-1}}_{i_1\ldots i_{p+q-1}}
\omega'_{\nu j_{p+1}\ldots j_{p+q-1}\alpha}{\omega_{j_1\ldots j_{p}}}^\nu
\right\}
x^\alpha\quad.
\cr}
\eqn\VIIiv
$$
We now state the theorem which gives all the linear GPS on simple 
Lie algebras:
\medskip
\noindent
{\bf Theorem 8.3}\quad
Let \g\ be a simple compact
\foot{
{\it Note.\quad}
The requirement of compactness is introduced to have a definite 
Killing--Cartan metric 
which then may be taken as the unit matrix; this is convenient to identify
upper and lower indices.
} 
algebra, and let $\omega$ and $\omega'$ be two 
non-trivial Lie algebra $(p+1)$- and $(q+1)$-cocycles obtained from the 
associated $p/2 +1$ and $q/2+1$ invariant symmetric tensors on $\g$.
Then, the associated skew-symmetric contravariant vector fields $\Omega$ and 
$\Omega'$ have zero SNB.

\noindent
{\it Proof.}\quad 
Since both $\Omega\in\Lambda^p(\g)\,,\,\Omega'\in\Lambda^q(\g)$ in \VIIiv\ have 
arbitrary even order, 
both terms have the same structure. It is thus sufficient to check that one of 
them is zero.
By \VIxiii\ and \VIxiv\
The first term gives
$$
\eqalign{&
\epsilon^{j_1\ldots j_{p+q-1}}_{i_1\ldots i_{p+q-1}}
\omega_{\nu j_1\ldots j_{p-1}\alpha}
{\omega'_{j_p\ldots j_{p+q-1}}}^\nu
\cr
=&
\epsilon^{j_1\ldots j_{p+q-1}}_{i_1\ldots i_{p+q-1}}
\epsilon^{l_1\ldots l_{p-1}}_{j_1\ldots j_{p-1}}
C^{s_1}_{\nu l_1}\ldots C^{s_{p/2}}_{l_{p-2}l_{p-1}}
k_{s_1\ldots s_{p/2}\alpha}
{\omega'_{j_p\ldots j_{p+q-1}}}^\nu
\cr
=&
(p-1)!
\epsilon^{l_1\ldots l_{p-1} j_p\ldots j_{p+q-1}}_{i_1\ldots i_{p+q-1}}
C^{s_1}_{\nu l_1}\ldots C^{s_{p/2}}_{l_{p-2}l_{p-1}}
k_{s_1\ldots s_{p/2}\alpha}
{\omega'_{j_p\ldots j_{p+q-1}}}^\nu
=0\quad,\cr}
\eqn\nuevath
$$
which is zero using \VIxii\ in the last equality {\it q.e.d.} 
\medskip

Since all the even order skew-symmetric multivector fields $\Omega$ associated 
with the odd order Lie algebra cocycles have zero SNB among themselves we find 
the
\medskip
\noindent
{\bf Corollary 8.1\quad}
Let $\g$ be a simple compact algebra, and let 
$k_{i_1\ldots i_m}$ be
a primitive invariant symmetric polynomial of order $m$.
Then, the tensor $\omega_{\rho l_2\ldots l_{2m-2} \sigma}$
$$
\Omega^{(2m-2)}={1\over (2m-2)!}{\omega_{l_1\ldots l_{2m-2}}}^\sigma
x_\sigma \partial^{l_1}\wedge\ldots\wedge \partial^{l_{2m-2}}\quad,
\eqn\fivetwo
$$
obtained from the cocycle \VIxiv, defines a linear GPS on $\g$. 
In particular (cf. \IVv) 
$$
\{x_{i_1},x_{i_2},\ldots,x_{i_{2m-2}}\}=
{\omega_{i_1\ldots i_{2m-2}}}^\sigma x_\sigma\quad,
\eqn\inpart
$$
where
${\omega_{i_1\ldots i_{2m-2}}}^\sigma$ are the `structure constants' 
defining the Lie algebra $(2m-1)$-cocycle.

Let $\Omega$ be as in eq. \VIIi\ and such that it defines a linear GPS.
Then, the following lemma follows:

\medskip
\noindent
{\bf Lemma 8.3\quad}
The operator
$\partial_\Omega:\wedge^q(\g)\to\wedge^{q+p-1}(\g)$ defined by
$$
(\partial_\Omega B)_{i_1\ldots i_{p+q-1}}={1\over p!}{1\over (q-1)!}
\epsilon^{j_1\ldots j_{p+q-1}}_{i_1\ldots i_{p+q-1}}
{\omega_{j_1\ldots j_p}}^\nu
B_{\nu j_{p+1}\ldots j_{p+q-1}}
\quad,
\eqn\coboper
$$
where $\Omega^{(2p)}$ is an even skew-symmetric contravariant tensor 
defining a linear GPS, is nilpotent, $\partial^2_\Omega=0$.

\noindent
{\it Proof.}\quad
>From the definition \coboper\ of $\partial_\Omega$,
$$
\eqalign{
(\partial^2_\Omega B)_{i_1\ldots i_{2p+q-2}}=
&
{1\over p!(p+q-2)!}
\epsilon^{j_1\ldots j_{2p+q-2}}_{i_1\ldots i_{2p+q-2}}
{\omega_{j_1\ldots j_p}}^\nu\;\cdot
\cr
\cdot
&
\left({1\over p!(q-1)!}
\epsilon^{k_1\ldots k_{p+q-1}}_{\nu j_{p+1} \ldots j_{2p+q-2}}
{\omega_{k_1\ldots k_p}}^\sigma
B_{\sigma k_{p+1}\ldots k_{p+q-1}}\right)
\cr
=
&
{1\over (p!)^2(p+q-2)!(q-1)!}
\sum_{s=1}^{p+q-1}(-1)^{s+1}
\epsilon^{j_1\ldots j_{2p+q-2}}_{i_1\ldots i_{2p+q-2}}
{\omega_{j_1\ldots j_p}}^{k_s}
\cr
\ &
\epsilon^{k_1\ldots \hat k_s \ldots k_{p+q-1}}_{j_{p+1} \ldots j_{2p+q-2}}
{\omega_{k_1\ldots k_p}}^\sigma 
B_{\sigma k_{p+1}\ldots k_{p+q-1}}\cr
=
&
{1\over (p!)^2(q-1)!}
\bigl[
\sum_{s=1}^p (-1)^{s+1}
\epsilon^{j_1\ldots j_p k_1 \ldots \hat k_s\ldots k_{p+1}\ldots k_{p+q-1}}
_{i_1 \ldots i_{2p+q-2}}
{\omega_{j_1\ldots j_p}}^{k_s}
\cr
+
&
\sum_{s=p+1}^{p+q-1}(-)^{s+1}
\epsilon^{j_1\ldots j_p k_1 \ldots k_{p}\ldots \hat k_s \ldots k_{p+q-1}}
_{i_1 \ldots i_{2p+q-2}}
{\omega_{j_1\ldots j_p}}^{k_s}\bigr]
{\omega_{k_1\ldots k_p}}^\sigma 
B_{\sigma k_{p+1}\ldots k_{p+q-1}} 
\cr
=
&
{p\over (p!)^2(q-1)!}
\epsilon^{j_1\ldots j_p k_1 \ldots k_{p+q-2}}_{i_1 \ldots i_{2p+q-2}}
{\omega_{j_1\ldots j_p}}^{\rho}
{\omega_{\rho k_1\ldots k_{p-1}}}^\sigma
B_{\sigma k_{p}\ldots k_{p+q-2}}
\cr
+
&
{(q-1)\over (p!)^2(q-1)!}
\epsilon^{j_1\ldots j_p k_1 \ldots k_{p+q-2}}_{i_1 \ldots i_{2p+q-2}}
{\omega_{j_1\ldots j_p}}^{\rho}
{\omega_{k_1\ldots k_{p}}}^\sigma
B_{\sigma \rho k_{p+1}\ldots k_{p+q-2}}=0\quad,
\cr}
\eqn\longproof
$$
since in the last equality the first term is zero on account of \nuevath, 
and the second one vanishes 
since $p$ is even and $B$ is skew-symmetric in $(\rho,\sigma)$, {\it q.e.d.}

In view of the above Lemma and eq. \VIIiv, Theorem 8.3 has the following
\medskip
\noindent
{\bf Corollary 8.2\quad}
If $\Omega$ and $\Omega'$ (of even order $p$ and $p'$) define two linear GPS, 
their SNB may be written as
$$
[\Omega,\Omega']=\partial_\Omega\Omega'+\partial_{\Omega'}\Omega\quad.
$$
In particular, $\partial_\Omega\Omega=0$ since $\Omega$ is a cocycle for 
$\partial_\Omega$.

Let us remark that the linear GPS given by the Lie algebra cocycles provide 
explicit examples of a non-decomposable (\ie, not given by the skew-symmetric 
product of single vectors) GPS.
In contrast, and as conjectured in \refmark{\ChaTak}, it has been recently shown 
\refmark{\AG} that all Nambu-type PS are decomposable (see also 
\refmark{\DFST} for more details on this point).
As an example of our theory consider the GPS which may be constructed on 
$su(3)^*$ defined by
$$
\Lambda^{(4)}={1\over 4!}{\omega_{i_1i_2i_3i_4}}^\sigma x_\sigma
\dd{x_{i_1}}\wedge\ldots\wedge\dd{x_{i_4}}
\quad,\quad
\omega_{\rho i_2i_3i_4 \sigma}:={1\over 2}
\epsilon^{j_2j_3j_4}_{i_2i_3i_4}d_{k_1k_2\sigma}C^{k_1}_{\rho j_2}
C^{k_2}_{j_3j_4}\quad,
\eqn\fivethree
$$
where the $d_{ijk}$ are the constants appearing in the anticommutator of the 
Gell--Mann $\lambda_i$ matrices,
$\{\lambda_i,\lambda_j\}={4\over 3}\delta_{ij}1_3+2d_{ijk}\lambda_k$.
It may be checked explicitly that
$[\Lambda^{(4)},\Lambda^{(4)}]=0$ (see \refmark{\APPBJPA} for details). 
Other examples may be given similarly.

\chapter{Conclusions}

We have established in this paper the mathematical basis of a new type of 
generalized Poisson structures.
>From a physical point of view, a more detailed investigation of the generalized 
Hamiltonian dynamics presented here and of its possible applications is needed;
clearly, one would like to have more examples besides the simple one provided 
in Sec. 6.
>From a mathematical point of view, the linear GPS are also interesting since, 
when applied to the case of the simple Lie algebras as in the example above, 
they provide the equivalent of the higher order Lie algebras \refmark{\HIGHER} 
which can be defined on any simple Lie algebra associated with its non trivial 
cohomology groups. This produces a set of examples (in fact, infinitely many of
them: $l$ GPS for {\it each} simple Lie algebra of rank $l$, of which the 
first one is the standard Lie-Poisson structure \IVv\ given by the structure 
constants
\foot{
This is in sharp contrast with the Nambu case, where only the
structure constants of $sl(2,\C)$ may serve as Nambu tensors \refmark{\ChaTak}
since those of the other simple algebras do not satisfy the `fundamental 
identity' 
(see also the Remark 1 in \refmark{\Ta}).
An interesting question is whether the set of linear GPS based 
on higher-order Lie algebras 
and those of the type discussed in the example of Sec. 6 (which trivially 
satisfy the GJI) constitute all the possible linear GPS.})
which illustrate our linear GPS in a non-trivial way.
The corresponding higher order simple Lie algebras \refmark{\HIGHER} 
are in turn special cases of the strongly homotopy 
Lie algebras (see \refmark{\LAST} and references therein) 
which have been found relevant in 
closed string theory (see, \eg, \refmark{\WZ,\ZIE}) and in connection with the 
Batalin-Vilkovisky antibracket.
Nevertheless, more work is needed to see whether the proposed GPS, very 
appealing by its 
geometrical contents, have also some direct physical applications.

We shall conclude here by saying that our analysis could be extended to
Lie superalgebras and super-Poisson structures in general by using an 
appropriate graded version of the SNB.
To this aim, one would need first a theory of skew graded multivector algebras,
which up to our knowledge is lacking \refmark{\AIPP}.
All these are directions for further research.

\bigskip
\noindent
{\bf Acknowledgements.\quad}
This research has been partially supported by the CICYT (Spain) under grant 
AEN93--187. 
A. P. wishes to thank the Vicerectorate of research of Valencia University for 
making his stay in Valencia possible;
J.C.P.B. wishes to acknowledge an FPI grant from the Spanish Ministry of 
Education and Science and the CSIC.
The authors whish to thank J. Stasheff for helpful comments on the manuscript.

\bigskip

\refout

\end